\documentclass[]{revtex4-2}

\usepackage{dsfont}
\usepackage{graphicx}
\usepackage{dcolumn}
\usepackage{bm}
\usepackage[utf8]{inputenc}
\usepackage{amsmath}
\usepackage {mathrsfs}

\begin{document}

\preprint{APS/123-QED}
\title{A Lagrangian with $E_8\times E_8$ symmetry for the standard model and pre-gravitation\ I.\\
{\it - The bosonic Lagrangian, and a theoretical derivation of the weak mixing angle -}}

\author{Sherry Raj}
 \email{sherry$_{\rm r}$@ph.iitr.ac.in}
\affiliation{ Indian Institute of Technology Roorkee\\
Roorkee, Uttarakhand 247667,  India}

\author{Tejinder P. Singh}
 \email{tpsingh@tifr.res.in \quad Address from January 1, 2023 : IUCAA, Pune}
\affiliation{Tata Institute of Fundamental Research,\\
Homi Bhabha Road, Mumbai 400005, India}





\bigskip

\bigskip

\begin{abstract}

\noindent Building on earlier work, we propose an elementary Lagrangian for the unification of the standard model with pre-gravitation, assumed to have an unbroken $E_8 \times E_8$ symmetry. The Lagrangian is patterned after the kinetic energy of a free particle in Newtonian dynamics, generalising it to the matrix-valued Lagrangian dynamics of a 2-brane on a split bioctonionic space. Symmetry breaking gives rise to the standard model quarks and leptons of three generations along with the known gauge interactions, and a novel right-handed counterpart $SU(3)_{grav}\times SU(2)_R \times U(1)_g$ identified as pre-gravitation. The goal of the present series of papers is to investigate if this pre-quantum, pre-spacetime matrix-valued Lagrangian dynamics can lead to the emergence of quantum field theory for the standard model, and classical general relativity, possibly with some additional corrections. In this paper we work out the bosonic part of the Lagrangian, and show how the anticipated 32 gauge bosons are recovered from the elementary Lagrangian. As an application, we show that the asymptotic low energy  weak mixing angle $\theta_W$ is given by a solution of the equation $1 = (1/2)\sqrt{ \cos(\theta_{W}/2)}+ \sqrt {\sin(\theta_{W}/2)}$, yielding $\sin^2\theta_W \approx 0.24996$. 

\end{abstract}

\maketitle

\section{Introduction}
\noindent When one takes the square-root of the Klein-Gordon equation to arrive at the spinorial equation for the electron, doing so does not involve any change in energy scale. Rather, it gives us a more precise formalism to describe the electron, one that agrees with experiment. In the same spirit, one can also consider taking the square-root of Minkowski space-time, so as to arrive at Penrose's twistor space, a spinor description of Minkowski spacetime. Again, no change of energy scale is involved, but one can try describing quantum dynamics of elementary particles on this spinor spacetime labelled by complex numbers, and look for possible advantages. In particular, we find that when complex numbers are replaced by the number system known as the octonions, the symmetries of the octonionic space reveal those of the standard model and of pre-gravitation \cite{priyank}. The algebra of the octonions shows evidence of fixing the values of  (at least some of) the parameters of the standard model \cite{Singhfsc, vvs}. The equivalent Minkowski spacetime is now ten dimensional, and the idea is that the four fundamental forces are the geometry of this space-time, and the associated pre-quantum dynamics is given by the theory of trace dynamics. It is important to emphasize that this new formalism is playing out already at the energies at which the standard model has been tested, and we have some understanding as to how the classical unverse, and quantum field theory on 4D classical spacetime, is recovered from the underlying pre-quantum spinor spacetime dynamics on octonionic twistor space \cite{Singhreview}. Fig. 1 below attempts to sketch this general idea.
\begin{figure}[ht]
    \includegraphics[width = 8cm]{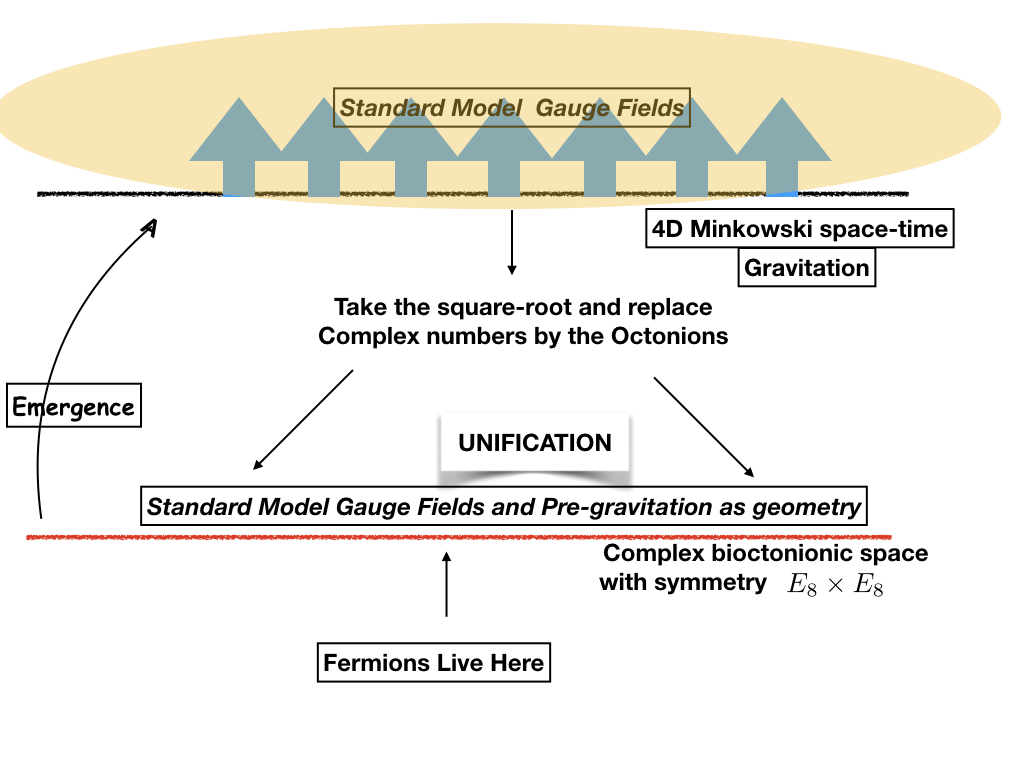}
    \caption{Fermions and bosons on a spinor spacetime labelled by the octonions \cite{singhessay}}
\end{figure}

In a recent work \cite{priyank} we have proposed the unbroken $E_8 \times E_8$ symmetry as the symmetry for unification of the standard model and pre-gravitation. Amongst many noteworthy features, one is that the 248 fundamental rep of $E_8$ is also its adjoint rep. This is appropriate for our theory in which a priori the fundamental degrees of freedom are neither bosonic nor fermionic, as will see below. The two $E_8$ branch as follows, after the $E_8 \times E_8$ symmetry is broken:
\begin{align}
& E_8 \rightarrow SU(3)_{Eucsp} \times E_6 \rightarrow SU(3)_{Eucsp} \times SU(3)_{genL} \times SU(3)_{color} \times SU(2)_L \times U(1)_Y \nonumber \\
& E_8 \rightarrow SU(3)_{spacetime} \times E_6 \rightarrow SU(3)_{spacetime} \times SU(3)_{genR} \times SU(3)_{grav} \times SU(2)_R \times U(1)_g
\label{branching}
\end{align}
The branching naturally induces separation of space-time-matter into space-time and matter (see Lagrangian below). The first $E_8$ branches into $SU(3)_{Eucsp} \times E_6$, and the 8D real rep of this $SU(3)$ is mapped on to an 8D vector space, this being the eight directions of an octonion $O$ (one real direction and seven imaginary). The absolute squared magnitude of the octonion has Euclidean signature and this space is equivalent to 10D Euclidean space $SO(10)$. The accompanying $E_6$ describes three generations of left-handed  quarks and leptons of the standard model, and their anti-particles, and twelve gauge bosons of the standard model $SU(3)_c \times SU(2)_L\times U(1)_Y$ and four Higgs boson degrees of freedom (most likely a charged Higgs).

The second $E_8$ branches into $SU(3)_{spacetime}\times E_6$, with the 8D real rep of $SU(3)_{spacetime}$ this time mapped to the split octonion part $O'$ (please see below for details) of a split bioctonion $O\oplus \omega O'$ ($\omega$ being the split complex number). The absolute squared magnitude of $\omega O'$ has Lorentzian signature and this octonionic space corresponds to 10D spacetime $SO(1,9)$. The associated $E_6$ describes three generations of right-handed quarks and leptons of the standard model (including sterile neutrinos), and their antiparticles, and twelve newly introduced gauge bosons describing the pre-gravitational symmetry $SU(3)_{grav}\times SU(2)_R \times U(1)_g$ and the standard model Higgs.

Between them, the two $E_8$, after symmetry breaking, give a picture akin to `fibre bundle describing gauge fields on spacetime'. Only, now the fibre bundle as well as spacetime have higher dimensions, and are defined on a bioctonionic spinor space. The endeavour is to build on earlier preliminary work and construct a Lagrangian which describes the original symmetry and its breaking, and to recover the observed 4D universe with standard model fields living on it. As an application, in the present paper we deduce the low energy asymptotic value of the weak mixing angle. A key idea is that quantum systems (such as an electron in flight) in fact obey the unbroken $E_8 \times E_8$ symmetry, even at low energies, and the symmetry appears broken because of the measurements we choose to make. We will also see that it is not possible to determine the value of the mixing angle without first unifying the standard model with pre-gravitation; nor without working on a higher dimensional spinor spacetime. The same was seen earlier for the derivation of the low energy fine structure constant as well as the mass ratios.

An octonion is made of one real unit and seven imaginary units $e_i^2 = -1$ for i = 1, 2,..., 7. Just as for the quaternions, the imaginary units of an octonion anti-commute. The multiplication of octonions is prescribed by a  diagram known as the Fano plane (Fig. 2)
\begin{figure}[ht]
\centering
\includegraphics[width=7.5cm]{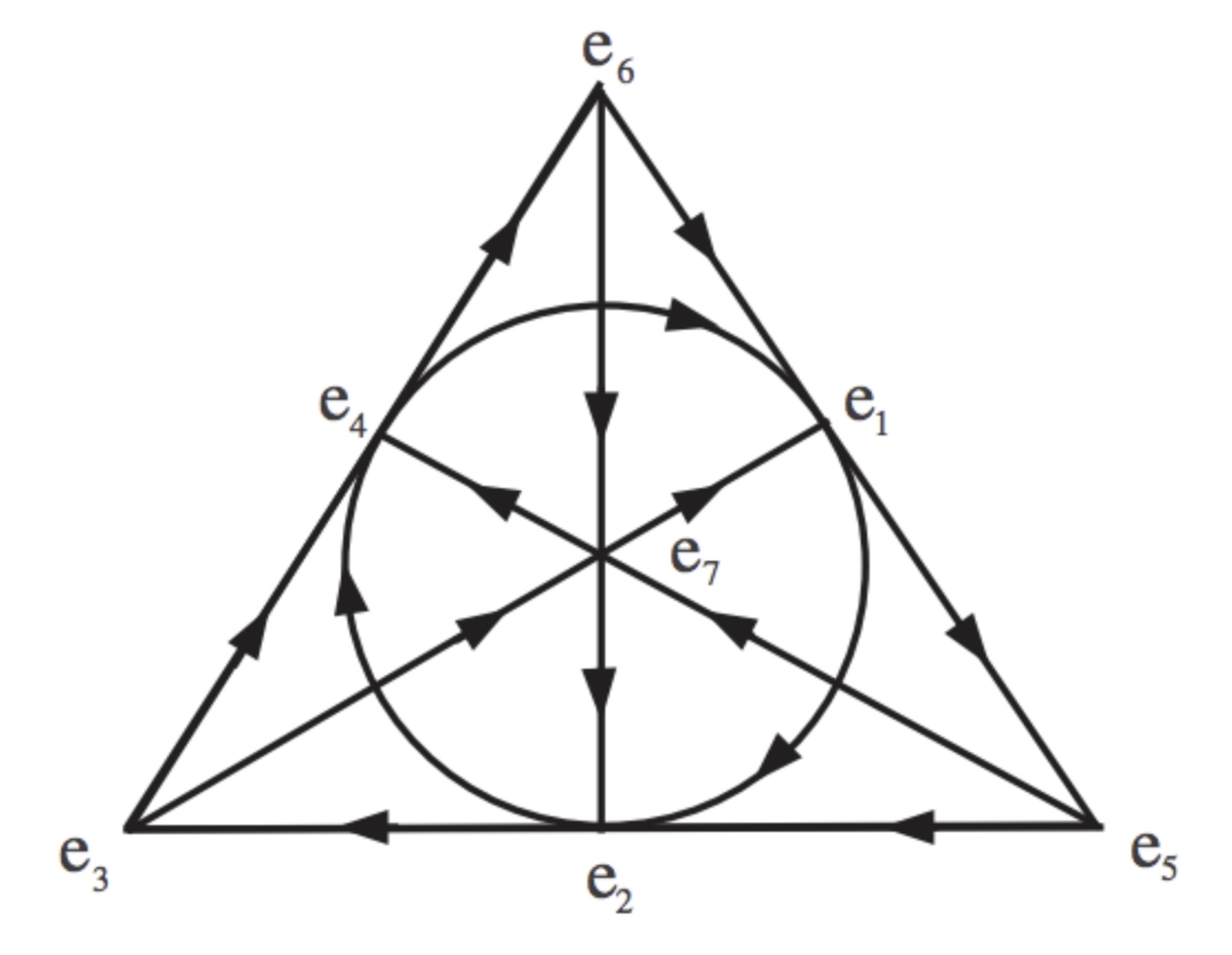}
\caption{The Fano plane gives rules for multiplying octonions}
\end{figure}
There are seven quaternionic subsets in the Fano plane, these are given by the three sides of the triangle, the three altitudes, and the circle. Multiplication of points lying along a quaternionic subset in cyclic order (as per the arrow) is given by $e_ie_j = e_k$, whereas $e_je_i = -e_k$. 

A split bioctonion $O_{sb}\equiv O \oplus \omega \tilde{O}$ is made from a pair of octonions as follows \cite{Vatsalya1}
\begin{equation}
O_{sb} =  (1, e_1, e_2, e_3, e_4, e_5, e_6, e_8) \oplus \omega(1, -e_1, -e_2, -e_3, -e_4, -e_5, -e_6, -e_8) 
\end{equation}
where $\tilde{O}$ is the octonionic conjugate of $O$, and $\omega$ is the equivalent of a split complex number, but made from the octonions themselves, as is seen in the context of the Clifford algebra $Cl(7)$ \cite{Vatsalya1}.  The fundamental degree of freedom in the octonionic theory is a 2-brane, also referred to as an `atom of space-time-matter', or an aikyon. It is defined by a pair of matrices $\widetilde{Q}_1$ and $\widetilde{Q}_2$ which have complex Grassmann numbers as their entries. Each of the two matrices has sixteen matrix-valued coordinate components, one for each of the sixteen  directions of the above bioctonion. The action principle for the 2-brane is given by \cite{Singhreview, q1q2uni}
\begin{equation}
\frac{S}{\hbar} = \int \frac{d\tau}{\tau_{Pl}}\; {\cal L} \qquad  ; \qquad\mathcal{L} =  \frac{L_p^2}{2L^2} Tr \biggl[\biggr.   \dot{\widetilde{Q}}_{1}^{\dagger}\;  \dot {\widetilde{Q}}_{2} \biggr]
\label{acnbasis}
\end{equation}
where $\tau$ is Connes time \cite{chamsnontech} and $L^2$ is the fundamental area of the 2-brane, in units of Planck area $L_p^2$. This is the only parameter in the theory to begin with, and its value is also fixed by the octonion algebra. Thus there are no free or fine tuned constants in the theory. Coupling constants and all other parameters of the standard model should emerge upon symmetry breaking, from the octonion algebra / geometry. We note that this Lagrangian is scale-invariant and assumed invariant under an unbroken $E_8\times E_8$ symmetry. Scale invariance is broken when this symmetry breaks, which is also Left-Right symmetry breaking as we will see below.

This matrix-valued Lagrangian dynamics is not to be quantised; rather it is a pre-quantum dynamics from which quantum field theory is emergent. This is as per the theory of trace dynamics \cite{Adler:94, AdlerMillard:1996, Adler:04}. Additionally, our theory is also a pre-spacetine dynamics, from which 4D curved spacetime and classical macroscopic objects are emergent \cite{Singhreview}. Therefore, gravitation, as well as quantum theory, are regarded as emergent phenomena.

An atom of space-time-matter is an elementary particle, say an electron, along with all the fields it produces. Of course the particle is fermionic and the produced fields are bosonic, but at the level of $\widetilde{Q}_1$ and $\widetilde{Q}_2$ one does not make a distinction between the bosonic aspect and the fermionic aspect. That distinction is introduced by noting that a Grassmann number can always be written as the sum of a Grassmann even part and a Grassmann odd part. Hence a matrix $\widetilde{Q}$ can be written as a sum of two matrices: $Q = Q_B + Q_F$, with the first one made from even-grade Grassmann numbers and called bosonic part, and the second one made from odd-grade Grassmann numbers and called fermionic.

Our universe is fundamentally made of enormously many aikyons, each having its own copy of $E_8\times E_8$ symmetry. Entanglement of sufficiently many aikyons localises their fermionic aspect, giving rise to macroscopic objects in the universe, and the emergence of 4D curved spacetime and classical gauge fields. Quantum degrees of freedom, i.e. those aikyons which are not critically entangled, continue to obey $E_8 \times E_8$ symmetry and are described precisely by the Lagrangian dynamics outlined below, or approximately, by the laws of quantum field theory on the emergent 4D spacetime. In the latter formalism, we lose the explanation for the origin of the standard model symmetries and for values of its parameters, because the memory of the underlying $E_8\times E_8$ symmetry is lost. The emergence of the classical theory is facilitated by the spectral action principle \cite{Chams:1997, Connes2000} which relates the eigenvalues of the Dirac operator (on which the present dynamics is based) to the action principle of the classical theory (gravitation and gauge fields and their matter sources).

Hence we define bosonic and fermionic degrees of freedom as follows
\begin{equation}  
\label{Qs0}
 \dot{\widetilde{Q}}_1^\dagger \equiv \dot{\widetilde{Q}}_B^\dagger +\frac{L_p^2}{L^2}\beta_1 \dot{\widetilde{Q}}_F^\dagger\ ; \qquad
 \dot{\tilde{Q}}_2\equiv \dot{\widetilde{Q}}_B + \frac{L_p^2}{L^2}\beta _2  \dot{\widetilde{Q}}_F  
       \end{equation}
       and use this expansion in the Lagrangian. Here $\beta_1$ and $\beta_2$ are two unequal odd-grade Grassmann numbers introduced so as to keep the Lagrangian bosonic, and the constant $L_p^2/L^2$ is essential for obtaining the correct classical limit of the theory. The bosonic matrix and the fermionic matrix each have sixteen components over the sixteen directions of the bioctonion, and the Yang-Mills coupling constant $\alpha$ is now introduced as follows 
  \begin{equation} 
\dot{\widetilde{Q}}_B= \frac{1}{L}(i\alpha q_B +L\dot{q}_B)\ ; \qquad
\dot{\widetilde{Q}}_F= \frac{1}{L}(i\alpha q_F +L\dot{q}_F)
\end{equation}     
The introduction of $\alpha / L$ is an important step in the theory; it breaks the scale invariance of the original Lagrangian (\ref{acnbasis}). The introduction of $L$ is essential for matching dimension of the undotted terms with that of the dotted terms. The dotted terms are defined only over the conjugated octonionic part $\tilde O$ of the split bioctonion and hence have eight matrix-valued coordinate components; these are related to pre-gravitation. The undotted parts are defined over the part ${O}$ of the split bioctonion and hence also have eight matrix-valued components; these are related to the standard model $SU(3)_c \times SU(2)_L \times U(1)_Y$. Pre-gravitation is $SU(3)_{grav} \times SU(2)_R \times U(1)_g$. In the split bioctonion, since the part $\omega \tilde{O}$ is interpreted as the parity reversal of $\tilde{O}$, and since $SU(2)_L$ and $SU(2)_R$ are chiral symmetries, the introduction of the standard model and of pre-gravitation breaks Left-Right symmetry and also breaks $E_8 \times E_8$ symmetry. The breaking of scale invariance also helps understand why general relativity obeys equivalence principle, whereas standard model forces do not; the latter depend on $L$.

In terms of these variables, the Lagrangian can be written as
\begin{equation}
\label{lagexp}
\qquad\mathcal{L} =  \frac{L_p^2}{2L^2} Tr \biggl[\biggr.   \left\{\frac{1}{L}(i\alpha q_B +L\dot{q}_B)^\dagger+ \frac{L_p^2}{L^2}\beta_1 . \frac{1}{L}(i\alpha q_F +L\dot{q}_F)^\dagger  \right\}\times \left\{\frac{1}{L}(i\alpha q_B +L\dot{q}_B)+ \frac{L_p^2}{L^2}\beta_2 . \frac{1}{L}(i\alpha q_F +L\dot{q}_F)  \right\}\biggr]
\end{equation}
When the brackets are opened out, there are 1024 terms. Another useful way to write the Lagrangian is to define the dynamical variables
\begin{equation}
q^\dagger_1 = q_B^\dagger + \frac{L_P^2}{L^2} \beta_1 q^\dagger_F \qquad; \qquad q_2 = q_B + \frac{L_P^2}{L^2} \beta_2 q_F
\end{equation}
which gives
\begin{equation}
\begin{aligned}
{\cal L} &=  \frac{L_P^2}{2L^2} \; Tr \left[ \left(\dot{q}_1^\dagger + \frac{i\alpha}{L} q_1^\dagger \right) \times \left(\dot{q}_2 + \frac{i\alpha}{L} q_2 \right)\right]\\
&= \frac{L_P^2}{2L^2} \; Tr \left[  \dot{q}_1^\dagger \dot{q}_2 - \frac{\alpha^2}{L^2} q_1^\dagger q_2  + \frac{i\alpha}{L} q_1^\dagger \dot{q}_2  + \frac{i\alpha}{L} \dot{q}_1^\dagger q_2 \right]
\end{aligned}
\label{lagba}
\end{equation}
We can  expand each of these four terms inside of the trace Lagrangian by using the definitions of $q_1$ and $q_2$:
\begin{equation}
\begin{aligned}
\dot{q}_1^\dagger \dot{q}_2 & =  \dot{q}_{B}^\dagger \dot{q}_{B}  + \frac{L_P^2}{L^2} \dot{q}_B^\dagger \beta_2 \dot{q}_F + \frac{L_P^2}{L^2} \beta_1 \dot{q}_F^\dagger \dot{q}_B + \frac{L_P^4}{L^4} \beta_1 \dot{q}_F^\dagger \beta_2 \dot{q}_F \\
q_1^\dagger q_2 & = q_B^\dagger {q}_B + \frac{L_P^2}{L^2} q_B^\dagger \beta_2 {q}_F + \frac{L_P^2}{L^2} \beta_1 q_F^\dagger {q}_B + \frac{L_P^4}{L^4}\beta_1 q_F^\dagger  \beta_2 {q}_F \\
q_1^\dagger \dot{q}_2 & =   q_B^\dagger \dot{q}_B + \frac{L_P^2}{L^2} q_B^\dagger \beta_2 \dot{q}_F + \frac{L_P^2}{L^2} \beta_1 q_F^\dagger \dot{q}_B + \frac{L_P^4}{L^4}\beta_1 q_F^\dagger  \beta_2 \dot{q}_F \\
\dot{q}_1^\dagger q_2 & = \dot{q}_{B}^\dagger {q}_{B}  + \frac{L_P^2}{L^2} \dot{q}_B^\dagger \beta_2 {q}_F + \frac{L_P^2}{L^2} \beta_1 \dot{q}_F^\dagger {q}_B + \frac{L_P^4}{L^4} \beta_1 \dot{q}_F^\dagger \beta_2 {q}_F\\
\end{aligned}
\label{subequ}
\end{equation}
This form of the Lagrangian displays terms for the bosonic variables (gauge fields), the interaction part of the action, and terms bilinear in fermionic varaibles.
The goal of the present series of papers is to investigate if this Lagrangian, expanded over the split bioctonionic space, describes the standard model as well as general relativity, possibly with some additional departures from currently known theory. In the present paper, starting with the next section, we write out in detail the bosonic part of the Lagrangian, and interpret it.

The equations of motion in this Lagrangian dynamics can be worked out from (\ref{lagba}) and are given by \cite{q1q2uni}
\begin{equation}
\ddot {q}_1^\dagger = - \frac{\alpha^2}{L^2} \; {q}_1^\dagger \; ; \qquad \ddot {q}_2 = - \frac{\alpha^2}{L^2} \; {q}_2
\end{equation}
and are discussed in a little more detail in \cite{q1q2uni}. Different particles are different vibrations of the 2-brane obeying these equations. In particular, the Dirac equation for three fermion generations arises as the eigenvalue equation for the constant matrices arising in the solution to these oscillator equations. Its symmetry group in ten dimensions is in fact $E_6$ which is also the automorphism group of the complexified exceptional Jordan algebra \cite{Dray1, Dray2}. When restricted to the Hermitean sector, we get the exceptional Jordan algebra, whose automorphism group is $F_4$ and which solves the eigenvalue problem for given values of quantised electric charge, and reveals a derivation of mass ratios \cite{Singhfsc, vvs}. 

Note that in the Lagrangian (\ref{lagba}) the cross-terms, i.e. the ones proportional to $i\alpha$, do not contribute to the equations of motion, with their contribution cancelling out. These cross terms form a total time derivative, and hence their time integral gives a boundary term, but there is no reason why the boundary term should vanish. Thus its time dependence is determined by the equations of motion. Remarkably, as we will see below, these cross-terms describe the $SU(2)_L\times SU(2)_R$  symmetry of electroweak - pregravitation sector, suggesting that the  sector $SU(3)_c \times SU(3)_{grav}$ [which comes from the fully dotted first two terms] determines the weak-pregrav sector, as if there is a gauge-gravity duality, with $SU(3)_{grav} $ actually belonging to gauge sector, and the weak force belonging to the gravity sector.

In our theory, the $E_8 \times E_8$ symmetry breaking is also left-right symmetry breaking, and it is clearly also the electroweak symmetry breaking. Thus we have the inescapable conclusion that prior to the EW breaking quantum gravity and unification come in play, and therefore that a de Sitter like scale-invariant cosmic expansion resets the Planck scale to the EW scale. This partly solves the hierarchy problem (effective Planck scale = EW scale), though of course it remains to be understood why the Higgs has the particular mass it does. Furthermore, since we have noted from the outset that description of matter fields on a spinor spacetime does not involve going to higher energies, we are compelled to note that the said symmetry breaking must be manifest even at currently studied low energies. This can be understood if the symmetry breaking process is the quantum-to-classical transition, which is precipitated whenever sufficiently many atoms of space-time-matter are entangled, resulting in the localisation of fermions and emergence of 4D classical spacetime. Quantum systems do not live in 4D spacetime; they live in the space having $E_8 \times E_8$ symmetry, at all energies. Therefore the (non-commuting) extra dimensions are never to be compactified. Only classical systems live in 4D (compactification without compactification) and arrive here dynamically.  Energy scale is not the key, but the degree of entanglement is. In the very early universe the energy scale is important, but only indirectly so. When the very early universe undergoing a de Sitter like scale-invariant expansion becomes cool enough to allow critical entanglement for the quantum-to-classical transition to take place, the $E_8 \times E_8$ symmetry is broken. What remains to be proved is that the freeze-out happens somewhere between 100 GeV and a TeV,  and to find the exact value of this transition energy scale.

\section{The bosonic Lagrangian}

Our Lagrangian is over a sixteen dimensional split bi-octonionic space ($O \oplus \omega \tilde{O} $)
\begin{equation}   \label{lagr}
 \mathscr{L}= \frac{L_p^2}{2L^2} Tr( \dot{\widetilde{Q}}_1^{\dagger} \dot{\widetilde{Q}}_2)   
\end{equation}
where
\begin{equation}  
\label{Qs}
 \dot{\widetilde{Q}}_1^\dagger = \dot{\widetilde{Q}}_B^\dagger +\frac{L_p^2}{L^2}\beta_1 \dot{\widetilde{Q}}_F^\dagger\ ; \qquad
 \dot{\widetilde{Q}}_2= \dot{\widetilde{Q}}_B + \frac{L_p^2}{L^2}\beta _2  \dot{\widetilde{Q}}_F  
       \end{equation}
Each matrix has sixteen components, and  
\begin{equation} 
\dot{\widetilde{Q}}_B= \frac{1}{L}(i\alpha q_B +L\dot{q}_B)\ ; \qquad
\dot{\widetilde{Q}}_F= \frac{1}{L}(i\alpha q_F +L\dot{q}_F)
\end{equation}
The bosonic terms of the Lagrangian are:  
\begin{equation}
\mathscr{L}_{bosonic}=\frac{L_p^2}{L^4}\left(\alpha ^2q_B^\dagger q_B +L^2 \dot{q}_B^\dagger \dot{q}_B -i\alpha L  q_B^\dagger \dot{q}_B +i\alpha L \dot{q}_B^\dagger q_B\right) 
\label{bpart}
\end{equation}
and these four terms can respectively be expanded as follows:

The first of these terms has a coefficient ${L_p^2 \alpha^2 / L^4 }$ as per Eqn. (\ref{bpart}) above.
\begin{align}
 q_B^\dagger q_B = &(q_{B0}e_0 -q_{B1}^\dagger e_1-q_{B2}^\dagger e_2-q_{B3}^\dagger e_3-q_{B4}^\dagger e_4-q_{B5}^\dagger e_5-q_{B6}^\dagger e_6-q_{B8}^\dagger e_8)\times\nonumber \\
&(q_{B0}e_0 +q_{B1} e_1+q_{B2} e_2+q_{B3} e_3+q_{B4} e_4+q_{B5} e_5+q_{B6} e_6 + q_{B8} e_8)\nonumber \\
 = \nonumber\\
  & q_{B0}^2 +q_{B1}^\dagger q_{B1} +q_{B2}^\dagger q_{B2} +q_{B3}^\dagger q_{B3} +q_{B4}^\dagger q_{B4} +q_{B5}^\dagger q_{B5} +q_{B6}^\dagger q_{B6} +q_{B8}^\dagger q_{B8}\nonumber\\
  -\ & (q_{B1}^\dagger e_1+q_{B2}^\dagger e_2+q_{B3}^\dagger e_3+q_{B4}^\dagger e_4+q_{B5}^\dagger e_5+q_{B6}^\dagger e_6+q_{B8}^\dagger)q_{B0}\nonumber\\
 +\ & q_{B0}(q_{B1} e_1 +q_{B2} e_2 +q_{B3} e_3 +q_{B4} e_4 +q_{B5} e_5 +q_{B6} e_6 +q_{B8} e_8)\nonumber\\
+\ & (-q_{B2}^\dagger q_{B4} +q_{B4}^\dagger q_{B2} -q_{B3}^\dagger q_{B8}+ q_{B8}^\dagger q_{B3} -q_{B5}^\dagger q_{B6} +q_{B6}^\dagger q_{B5}) e_1\nonumber\\\
+\ & (-q_{B4}^\dagger q_{B1} +q_{B1}^\dagger q_{B4} -q_{B3}^\dagger q_{B5} +q_{B5}^\dagger q_{B3} -q_{B6}^\dagger q_{B8} +q_{B8}^\dagger q_{B6})e_2\nonumber\\
+\ &(-q_{B4}^\dagger q_{B6} +q_{B6}^\dagger q_{B4} -q_{B5}^\dagger q_{B2} +q_{B2}^\dagger q_{B5} -q_{B8}^\dagger q_{B1} +q_{B1}^\dagger q_{B8})e_3\nonumber\\
+\ & (-q_{B1}^\dagger q_{B2} +q_{B2}^\dagger q_{B1} -q_{B6}^\dagger q_{B3} +q_{B3}^\dagger q_{B6} -q_{B5}^\dagger q_{B8} +q_{B8}^\dagger q_{B5})e_4\nonumber\\
+\ &(-q_{B2}^\dagger q_{B3} +q_{B3}^\dagger q_{B2} -q_{B6}^\dagger q_{B1} +q_{B1}^\dagger q_{B6} -q_{B8}^\dagger q_{B4} +q_{B4}^\dagger q_{B8})e_5\nonumber\\
+\ &(-q_{B1}^\dagger q_{B5} +q_{B5}^\dagger q_{B1} -q_{B8}^\dagger q_{B2} +q_{B2}^\dagger q_{B8} -q_{B3}^\dagger q_{B4} +q_{B4}^\dagger q_{B3})e_6\nonumber\\
+\ &(-q_{B1}^\dagger q_{B3} +q_{B3}^\dagger q_{B1} -q_{B2}^\dagger q_{B6} +q_{B6}^\dagger q_{B2} -q_{B4}^\dagger q_{B5} +q_{B5}^\dagger q_{B4})e_8 
\label{qbqb}
\end{align}
This term has a coefficient ${L_p^2 / L^2 }$ as per Eqn. (\ref{bpart}) above.
\begin{align}
 \dot{q}_B^\dagger \dot{q}_B = &(\dot q_{B0}e_0 -\dot q_{B1}^\dagger e_1-\dot q_{B2}^\dagger e_2-\dot q_{B3}^\dagger e_3-\dot q_{B4}^\dagger e_4-\dot q_{B5}^\dagger e_5-\dot q_{B6}^\dagger e_6-\dot q_{B8}^\dagger e_8)\times\nonumber \\
&(\dot q_{B0}e_0 +\dot q_{B1} e_1+\dot q_{B2} e_2+\dot q_{B3} e_3+\dot q_{B4} e_4+\dot q_{B5} e_5+\dot q_{B6} e_6 + \dot q_{B8} e_8)\nonumber \\
 = \nonumber\\
& \dot{q}_{B0}^2 +\dot{q}_{B1}^\dagger \dot{q}_{B1} +\dot{q}_{B2}^\dagger \dot{q}_{B2} +\dot{q}_{B3}^\dagger \dot{q}_{B3} +\dot{q}_{B4}^\dagger \dot{q}_{B4} +\dot{q}_{B5}^\dagger \dot{q}_{B5} +\dot{q}_{B6}^\dagger \dot{q}_{B6} +\dot{q}_{B8}^\dagger \dot{q}_{B8}\nonumber\\ 
+\ &\dot{q}_{B0}(\dot{q}_{B1} e_1 +\dot{q}_{B2} e_2 +\dot{q}_{B3} e_3 +\dot{q}_{B4} e_4 +\dot{q}_{B5} e_5 +\dot{q}_{B6} e_6 +\dot{q}_{B8} e_8)\nonumber\\
+\ &(-\dot{q}_{B1}^\dagger e_1 -\dot{q}_{B2}^\dagger e_2 -\dot{q}_{B3}^\dagger e_3 -\dot{q}_{B4}^\dagger e_4 -\dot{q}_{B5}^\dagger e_5 -\dot{q}_{B6}^\dagger e_6 -\dot{q}_{B8}^\dagger e_8) \dot{q}_{B0}\nonumber\\
+\ &(-\dot{q}_{B2}^\dagger \dot{q}_{B4} +\dot{q}_{B4}^\dagger \dot{q}_{B2} -\dot{q}_{B3}^\dagger \dot{q}_{B8}+ \dot{q}_{B8}^\dagger \dot{q}_{B3} -\dot{q}_{B5}^\dagger \dot{q}_{B6} +\dot{q}_{B6}^\dagger \dot{q}_{B5}) e_1\nonumber\\
+\ &(-\dot{q}_{B4}^\dagger \dot{q}_{B1} +\dot{q}_{B1}^\dagger \dot{q}_{B4} -\dot{q}_{B3}^\dagger \dot{q}_{B5} +\dot{q}_{B5}^\dagger \dot{q}_{B3} -\dot{q}_{B6}^\dagger \dot{q}_{B8} +\ \dot{q}_{B8}^\dagger \dot{q}_{B6})e_2\nonumber\\
+\ &(-\dot{q}_{B4}^\dagger \dot{q}_{B6} +\dot{q}_{B6}^\dagger \dot{q}_{B4} -\dot{q}_{B5}^\dagger \dot{q}_{B2} +\dot{q}_{B5}^\dagger \dot{q}_{B2} -\dot{q}_{B8}^\dagger \dot{q}_{B1} +\dot{q}_{B1}^\dagger \dot{q}_{B8})e_3\nonumber\\
+\ &(-\dot{q}_{B1}^\dagger \dot{q}_{B2} +\dot{q}_{B2}^\dagger \dot{q}_{B1} -\dot{q}_{B4}^\dagger \dot{q}_{B6} +\dot{q}_{B6}^\dagger \dot{q}_{B3} -\dot{q}_{B5}^\dagger \dot{q}_{B8} +\dot{q}_{B8}^\dagger \dot{q}_{B5})e_4\nonumber\\
+\ &(-\dot{q}_{B2}^\dagger \dot{q}_{B3} +\dot{q}_{B3}^\dagger \ dot{q}_{B2} -\dot{q}_{B6}^\dagger \dot{q}_{B1} +\dot{q}_{B1}^\dagger \dot{q }_{B6} -\dot{q}_{B8}^\dagger \dot{q}_{B4} +\dot{q}_{B4}^\dagger \dot{q}_{B8})e_5\nonumber\\
+\ &(-\dot{q}_{B1}^\dagger \dot{q}_{B5} +\dot{q}_{B5}^\dagger \dot{q}_{B1} -\dot{q}_{B8}^\dagger \dot{q}_{B2} +\dot{q}_{B2}^\dagger \dot{q}_{B8} -\dot{q}_{B3}^\dagger \dot{q}_{B4} +\dot{q}_{B4}^\dagger \dot{q}_{B3})e_6\nonumber\\
+\ &(-\dot{q}_{B1}^\dagger \dot{q}_{B3} +\dot{q}_{B3}^\dagger \dot{q}_{B1} -\dot{q}_{B2}^\dagger \dot{q}_{B6} +\dot{q}_{B6}^\dagger \dot{q}_{B2} -\dot{q}_{B4}^\dagger \dot{q}_{B5} +\dot{q}_{B5}^\dagger \dot{q}_{B4})e_8 
\label{rhgluons}
\end{align}
These terms have a coefficient $-i{L_p^2 \alpha/ L^3 }$ as per Eqn. (\ref{bpart}) above.

\begin{align}
\left( q_B^\dagger \dot q_B - \dot q  _B^\dagger q_B\right)= & \omega (\dot{q}_{B0} q_{B0} -q_{B0} \dot{q}_{B0} -\dot{q}_{B1}^\dagger q_{B1} +q_{B1}^\dagger \dot{q}_{B1} -\dot{q}_{B2}^\dagger q_{B2} +q_{B2}^\dagger \dot{q}_{B2} -\dot{q}_{B3}^\dagger q_{B3} +q_{B3}^\dagger \dot{q}_{B3}\nonumber\\
-\ &\dot{q}_{B4}^\dagger q_{B4} +q_{B4}^\dagger \dot{q}_{B4} -\dot{q}_{B5}^\dagger q_{B5} +q_{B5}^\dagger \dot{q}_{B5} -\dot{q}_{B6}^\dagger q_{B6} +q_{B6}^\dagger \dot{q}_{B6} -\dot{q}_{B8}^\dagger q_{B8} +q_{B8}^\dagger \dot{q}_{B8})\nonumber\\
+\ &(\dot{q}_{B1}^\dagger q_{B0} -\dot{q}_{B4}^\dagger q_{B2} +\dot{q}_{B2}^\dagger q_{B4} -\dot{q}_{B6}^\dagger q_{B5} +\dot{q}_{B5}^\dagger q_{B6} -\dot{q}_{B8}^\dagger q_{B3} +\dot{q}_{B3}^\dagger q_{B8}   +q_{B1}^\dagger \dot{q}_{B0} +q_{B4}^\dagger \dot{q}_{B2} -q_{B2}^\dagger \dot{q}_{B4}\nonumber\\ & +q_{B6}^\dagger \dot{q}_{B5} -q_{B5}^\dagger \dot{q}_{B6} +q_{B8}^\dagger \dot{q}_{B3} -q_{B3}^\dagger \dot{q}_{B8})\omega e_1\nonumber\\
+\ &(\dot{q}_{B2}^\dagger q_{B0} -\dot{q}_{B1}^\dagger q_{B4} +\dot{q}_{B4}^\dagger q_{B1} -\dot{q}_{B5}^\dagger q_{B3} +\dot{q}_{B3}^\dagger q_{B5} -\dot{q}_{B8}^\dagger q_{B6}+\dot{q}_{B6}^\dagger q_{B8}   +q_{B2}^\dagger \dot{q}_{B0} +q_{B1}^\dagger \dot{q}_{B4} -q_{B4}^\dagger \dot{q}_{B1}\nonumber\\  & +q_{B5}^\dagger \dot{q}_{B3} -q_{B3}^\dagger \dot{q}_{B5} +q_{B8}^\dagger \dot{q}_{B6} -q_{B6}^\dagger \dot{q}_{B8})\omega e_2\nonumber\\
+\ &(\dot{q}_{B3}^\dagger q_{B0} -\dot{q}_{B1}^\dagger q_{B8} +\dot{q}_{B8}^\dagger  q_{B1} -\dot{q}_{B2}^\dagger q_{B5} +\dot{q}_{B5}^\dagger q_{B2} -\dot{q}_{B6}^\dagger q_{B4} +\dot{q}_{B4}^\dagger q_{B6}    +q_{B3}^\dagger \dot{q}_{B0} +q_{B1}^\dagger \dot{q}_{B8} -q_{B8}^\dagger  \dot{q}_{B1}\nonumber\\ &  +q_{B2}^\dagger \dot{q}_{B5} -q_{B5}^\dagger \dot{q}_{B2} +q_{B6}^\dagger \dot{q}_{B4} -q_{B4}^\dagger \dot{q}_{B6})\omega e_3\nonumber\\
+\ &(\dot{q}_{B4}^\dagger q_{B0} -\dot{q}_{B2}^\dagger q_{B1} +\dot{q}_{B1}^\dagger q_{B2} -\dot{q}_{B3}^\dagger q_{B6} +\dot{q}_{B6}^\dagger q_{B3} -\dot{q}_{B8}^\dagger q_{B5} +\dot{q}_{B5}^\dagger q_{B8}    +q_{B4}^\dagger \dot{q}_{B0} +q_{B2}^\dagger \dot{q}_{B1} -q_{B1}^\dagger \dot{q}_{B2}\nonumber\\ &  +q_{B3}^\dagger \dot{q}_{B6} -q_{B6}^\dagger \dot{q}_{B3} +q_{B8}^\dagger \dot{q}_{B5} -q_{B5}^\dagger \dot{q}_{B8})\omega e_4\nonumber\\
  +\ &(\dot{q}_{B5}^\dagger q_{B0} -\dot{q}_{B1}^\dagger q_{B6} +\dot{q}_{B6}^\dagger q_{B1} -\dot{q}_{B3}^\dagger q_{B2} +\dot{q}_{B2}^\dagger q_{B3} -\dot{q}_{B4}^\dagger q_{B8} +\dot{q}_{B8}^\dagger q_{B4}   +q_{B5}^\dagger \dot{q}_{B0} +q_{B1}^\dagger \dot{q}_{B6} -q_{B6}^\dagger \dot{q}_{B1}\nonumber\\ & +q_{B3}^\dagger \dot{q}_{B2} -q_{B2}^\dagger \dot{q}_{B3} +q_{B4}^\dagger \dot{q}_{B8} -q_{B8}^\dagger \dot{q}_{B4})\omega e_5\nonumber\\
+\ &(\dot{q}_{B6}^\dagger q_{B0} -\dot{q}_{B5}^\dagger q_{B1} +\dot{q}_{B1}^\dagger q_{B5} -\dot{q}_{B2}^\dagger q_{B8} +\dot{q}_{B8}^\dagger q_{B2} -\dot{q}_{B4}^\dagger q_{B3} +\dot{q}_{B3}^\dagger q_{B4}   +q_{B6}^\dagger \dot{q}_{B0} +q_{B5}^\dagger \dot{q}_{B1} -q_{B1}^\dagger \dot{q}_{B5}\nonumber\\ &  +q_{B2}^\dagger \dot{q}_{B8} -q_{B8}^\dagger \dot{q}_{B2} +q_{B4}^\dagger \dot{q}_{B3} -q_{B3}^\dagger \dot{q}_{B4})\omega e_6\nonumber\\
+\ &(\dot{q}_{B8}^\dagger q_{B0} -\dot{q}_{B3}^\dagger q_{B1} +\dot{q}_{B1}^\dagger q_{B3} -\dot{q}_{B6}^\dagger q_{B2} +\dot{q}_{B2}^\dagger q_{B6} -\dot{q}_{B5}^\dagger q_{B4} +\dot{q}_{B4}^\dagger q_{B5}  +q_{B8}^\dagger \dot{q}_{B0} +q_{B3}^\dagger \dot{q}_{B1} -q_{B1}^\dagger \dot{q}_{B3}\nonumber\\ &  +q_{B6}^\dagger \dot{q}_{B2} -q_{B2}^\dagger \dot{q}_{B6} +q_{B5}^\dagger \dot{q}_{B4} -q_{B4}^\dagger \dot{q}_{B5})\omega e_8
\label{weakgrav}
\end{align}


\subsection{Interpreting the terms of the bosonic Lagrangian}

\noindent We require a total of 32 bosons - 16 from the LH sector and 16 from the RH sector. We propose the following interpretation for the terms that have been explicitly written above and explain how we account for the 32 bosons:

\ 

1.\ \textbf{The Higgs boson}: We can identify two Higgs bosons ($Higgs_{LH}$ and $Higgs_{RH}$), one from each sector, as follows:

 \begin{equation} Higgs_{LH}=(L_p^2/L^4) \alpha ^2 q_{B0}^2 \; \qquad  Higgs_{RH}=(L_p^2/L^2) \dot{q}_{B0}^2
\end{equation}
Because these terms come as coefficients of the real direction $e_0$, we expect these bosons to be scalars. The additional terms associated with the Higgs are anticipated to come from the bifermionic part of the Lagrangian and give total of four bosonic degrees of freedom for each of the Higgs prior to symmetry breaking. This is in accordance with the proposed symmetry breaking of $E_8\times E_8$ \cite{priyank}. A detailed investigation of the Higgs is left for a later paper in this series where we will analyse the bifermion part of the Lagrangian.

2. \textbf{The photon}: In deriving the standard model from the algebra of the octonions, the $SU(3)$ subgroup of $G_2$ is identified with $SU(3)_{color}$. The ladder operators of the related Clifford algebra define a number operator.  In consistency with how the photon arises as the boson associated with the  $U(1)$ symmetry coming from the  number operator, we identify $U(1)_{em}$ from Eqn. (\ref{qbqb}) as the following expression:
\begin{equation}
 \frac{L_p^2}{L^4}\alpha ^2\left(q_{B1}^\dagger q_{B1} +q_{B2}^\dagger q_{B2} +q_{B3}^\dagger q_{B3} +q_{B4}^\dagger q_{B4} +q_{B5}^\dagger q_{B5} +q_{B6}^\dagger q_{B6} +q_{B8}^\dagger q_{B8}\right)
 \label{photon}
\end{equation}
The remaining terms in (\ref{qbqb}) define the eight gluons of QCD as we will see below, and hence the expanded terms of this equation define the nine bosons of the unbroken symmetry $SU(3)_c\times U(1)_{em}$ of the standard model. Thus the part $(L_P^2\alpha^2/L^4) q_B^\dagger q_B$ of the bosonic Lagrangian is fully accounted for; we recall that it is defined on the left-handed part of the bioctonion space.

3.\ \textbf{The $\mathbf{U(1)_{grav}}$ boson}: In a manner similar to the photon, a $U(1)_{grav}$ boson can be identified from Eqn. (\ref{rhgluons}), and is given by
\begin{equation}
    \frac{L_p^2}{L^2}\left(\dot{q}_{B1}^\dagger \dot{q}_{B1} +\dot{q}_{B2}^\dagger \dot{q}_{B2}+\dot{q}_{B3}^\dagger \dot{q}_{B3} +\dot{q}_{B4}^\dagger \dot{q}_{B4}+\dot{q}_{B5}^\dagger \dot{q}_{B5}+\dot{q}_{B6}^\dagger \dot{q}_{B6}+\dot{q}_{B8}^\dagger \dot{q}_{B8}\right)
    \end{equation}
    The role of this boson (sometimes referred to as the dark photon) in modifying general relativity, and its possible implications for cosmology, remain to be understood. 


4. \textbf{Gluons and gravi-gluons}:

We require the representations of the eight gluons to be obeying $SU(3)$ symmetry. In the language of the octonion algebra, the $SU(3)$ generators are given by Furey \cite{f1,f3} as follows (ignoring the overall multiplicative factor in the front):
 \label{generatorsso3}
\begin{align*}  
\Lambda_1f = e_1(e_5f)-e_3(e_4f)\\
\Lambda_2f= -e_1(e_4f)-e_3(e_5f)\\  
 \Lambda_3f= e_4(e_5f)-e_1(e_3f)\\
 \Lambda_4f= e_2(e_5f)+e_4(e_6f)\\
 \Lambda_5f=e_5(e_6f)-e_2(e_4f)\\
 \Lambda_6f=e_1(e_6f)+e_2(e_3f)\\ 
 \Lambda_7f=e_2(e_1f)+e_3(e_6f)\\
 \Lambda_8f=e_1(e_3f)+e_4(e_5f)-2e_2(e_6f)
  \end{align*}
Here $f$ is any octonion on which the generators act, and in our work we choose $f=1$ throughout, without loss of generality. Let us associate with each term $ e_i(e_j f)$ a term of the type:
\begin{equation}({q}_{Bi}^\dagger q_{Bj} -{q}_{Bj}^\dagger q_{Bi})e_k
\label{ansatz}
 \end{equation}
 for the left-handed sector, and the term
\begin{equation}(\dot q_{Bi}^\dagger \dot{q}_{Bj} -\dot q_{Bj}^\dagger \dot{q}_{Bi})e_k
 \end{equation}
for the right-hand sector, where $e_i e_j = e_k$. The spirit behind doing this is the same as in the general theory of relativity: introduce fields as the curvature / non-trivial geometry of flat spacetime. For us the analog of flat space-time is the complex bioctonionic space, and `curved space' is introduced through these matrix-valued dynamical variables, now representing pre-gravitation as well as standard model forces.

 Keeping this mapping in mind and setting $f=1$, we have the following representations for the {gluons}, coming from eight $SU(3)$ generators
\begin{align}
  \frac{Lp^2}{L^4} \alpha^2 [({q}_{B1}^\dagger q_{B5} -{q}_{B5}^\dagger q_{B1}) -(q_{B3}^\dagger {q}_{B4} -q_{B4}^\dagger {q}_{B3}) ]  e_6\nonumber\\
 \frac{Lp^2}{L^4} \alpha^2 [({q}_{B1}^\dagger q_{B4} -{q}_{B4}^\dagger q_{B1}) +(q_{B3}^\dagger {q}_{B5} -q_{B5}^\dagger {q}_{B3}) ]  e_2\nonumber\\
 \frac{Lp^2}{L^4} \alpha^2[({q}_{B4}^\dagger q_{B5} -{q}_{B5}^\dagger q_{B4}) -(q_{B1}^\dagger {q}_{B3} -q_{B3}^\dagger {q}_{B1}) ]  e_8\nonumber\\
\frac{Lp^2}{L^4} \alpha^2[({q}_{B4}^\dagger q_{B6} -{q}_{B6}^\dagger q_{B4}) +(q_{B2}^\dagger {q}_{B5} -q_{B5}^\dagger {q}_{B2}) ]  e_3\nonumber\\
 \frac{Lp^2}{L^4} \alpha^2[({q}_{B5}^\dagger q_{B6} -{q}_{B6}^\dagger q_{B5}) -(q_{B2}^\dagger {q}_{B4} -q_{B4}^\dagger {q}_{B2}) ]  e_1\nonumber\\
 \frac{Lp^2}{L^4} \alpha^2[({q}_{B2}^\dagger q_{B3} -{q}_{B3}^\dagger q_{B2}) +(q_{B1}^\dagger {q}_{B6} -q_{B6}^\dagger {q}_{B1}) ]  e_5\nonumber\\
 \frac{Lp^2}{L^4} \alpha^2[({q}_{B1}^\dagger q_{B2} -{q}_{B2}^\dagger q_{B1}) -(q_{B3}^\dagger {q}_{B6} -q_{B6}^\dagger {q}_{B3}) ]  e_4\nonumber\\
 \frac{Lp^2}{L^4} \alpha^2[({q}_{B1}^\dagger q_{B3} -{q}_{B3}^\dagger q_{B1}) + ({q}_{B4}^\dagger q_{B5} -{q}_{B5}^\dagger q_{B4}) -2(q_{B2}^\dagger {q}_{B6} -q_{B6}^\dagger {q}_{B2}) ]  e_8
 \label{su3genleft}
     \end{align}
     These expressions for the generators should be compared with the terms we have in Eqn. (\ref{qbqb}). The following terms in (\ref{qbqb}) are left out of the said comparison:
     \begin{align} \label{leftoutgluon}
     -\ & (q_{B1}^\dagger e_1+q_{B2}^\dagger e_2+q_{B3}^\dagger e_3+q_{B4}^\dagger e_4+q_{B5}^\dagger e_5+q_{B6}^\dagger e_6+q_{B8}^\dagger)q_{B0}\nonumber\\
 +\ & q_{B0}(q_{B1} e_1 +q_{B2} e_2 +q_{B3} e_3 +q_{B4} e_4 +q_{B5} e_5 +q_{B6} e_6 +q_{B8} e_8)
     \end{align}
The remaining terms of (\ref{qbqb}), apart from the photon, are compared with the forms given in (\ref{su3genleft}). We find that all the antisymmetric pairs of (\ref{su3genleft}) are present in (\ref{qbqb}) either exactly, or up to a sign, or obtainable by a relabelling of the $e_i$. The sign can easily be adjusted from the order of the two terms in (\ref{ansatz}). We can therefore  conclude that the gluons of $SU(3)_{color}$ symmetry are present in our proposed Lagrangian.

\bigskip

Similarly, we obtain the expressions coming from the generators of $SU(3)_{grav}$ for the gravi-gluons:   
\begin{align*}
 \frac{Lp^2}{L^2} [(\dot q_{B1}^\dagger \dot{q}_{B5}-\dot q_{B5}^\dagger \dot{q}_{B1}) -(\dot{q}_{B3}^\dagger \dot q_{B4} -\dot{q}_{B4}^\dagger \dot q_{B3})]  e_6,\\
 \frac{Lp^2}{L^2} [(\dot q_{B1}^\dagger \dot{q}_{B4}-\dot q_{B4}^\dagger \dot{q}_{B1}) +(\dot{q}_{B3}^\dagger \dot q_{B5} -\dot{q}_{B5}^\dagger \dot{} q_{B3})]  e_2,\\
 \frac{Lp^2}{L^2} [(\dot q_{B4}^\dagger \dot{q}_{B5}-\dot q_{B5}^\dagger \dot{q}_{B4}) -(\dot{q}_{B1}^\dagger \dot q_{B3} -\dot{q}_{B3}^\dagger \dot q_{B1})] e_8,\\
 \frac{Lp^2}{L^2} [(\dot q_{B4}^\dagger \dot{q}_{B6}-\dot q_{B6}^\dagger \dot{q}_{B4}) +(\dot{q}_{B2}^\dagger \dot q_{B5} -\dot{q}_{B5}^\dagger \dot q_{B2})]  e_3,\\
 \frac{Lp^2}{L^2} [(\dot q_{5}^\dagger \dot{q}_{B6}-\dot q_{B6}^\dagger \dot{q}_{B5}) -(\dot{q}_{B2}^\dagger \dot q_{B4} -\dot{q}_{B4}^\dagger \dot q_{B2})]  e_1,\\
 \frac{Lp^2}{L^2} [(\dot q_{B2}^\dagger \dot{q}_{B3}-\dot q_{B3}^\dagger \dot{q}_{B2}) +(\dot{q}_{B1}^\dagger \dot q_{B6} -\dot{q}_{B6}^\dagger \dot q_{B1})]  e_5,\\
 \frac{Lp^2}{L^2} [(\dot q_{B1}^\dagger \dot{q}_{B2}-\dot q_{B2}^\dagger \dot{q}_{B1}) +(\dot{q}_{B3}^\dagger \dot q_{B6} -\dot{q}_{B6}^\dagger \dot q_{B3})] e_4,\\
\frac{Lp^2}{L^2} [(\dot q_{B1}^\dagger \dot{q}_{B3}-\dot q_{B3}^\dagger \dot{q}_{B1})+(\dot q_{B4}^\dagger \dot{q}_{B5}-\dot q_{B5}^\dagger \dot{q}_{B4}) -2(\dot{q}_{B2}^\dagger \dot q_{B6} -\dot{q}_{B6}^\dagger \dot q_{B2})]  e_8
        \end{align*}
Like in the case of the gluons, these terms should be compared with the part (\ref{rhgluons}) of the bosonic Lagrangian above. Similar conclusions as for the case of the gluons hold. We thus have all the 8 reps for the gluons and the gravi-gluons each, as mapped from the SU(3) generators, and found in the Lagrangian. Furthermore, the left-over terms from (\ref{rhgluons}) are given by
\begin{align} \label{leftoutgravigluon}
     -\ & (\dot q_{B1}^\dagger e_1+\dot q_{B2}^\dagger e_2+\dot q_{B3}^\dagger e_3+\dot q_{B4}^\dagger e_4+ \dot q_{B5}^\dagger e_5+\dot q_{B6}^\dagger e_6+\dot q_{B8}^\dagger)\dot q_{B0}\nonumber\\
 +\ & \dot q_{B0}(\dot q_{B1} e_1 +\dot q_{B2} e_2 +\dot q_{B3} e_3 +\dot q_{B4} e_4 +\dot q_{B5} e_5 +\dot q_{B6} e_6 +\dot q_{B8} e_8)
     \end{align}
     The Lagrangian for the gluons of $SU(3)_{color}$ and the Lagrangian for the gravi-gluons of $SU(3)_{grav}$ exhibit an important symmetry: they can be mapped one to the other by the mapping $(\alpha/L) q_B \leftrightarrow \dot q_B$, suggesting this to be a gauge-gravity duality. This could also help understand why QCD is not chiral, even though in this octonionic approach the $SU(3)_{color}$ symmetry is constructed from the quarks of the left-handed sector. We note that $SU(3)_{grav}$ is constructed from quarks and leptons of the right-handed sector. The symmetry $SU(3)_{color}\times SU(3)_{grav}$ is the unbroken part of the $E_8 \times E_8$ symmetry, which we know is also a left-right symmetry. The actual force here is not QCD color but rather it is color-grav, described by $SU(3)_{color}\times SU(3)_{grav}$, and this force is not chiral. It is obeyed both by left-handed particles as well as right-handed particles. If we assume that the grav part is extremely weak at the nuclear scale, compared to the color part, the force will appear as if it is only $SU(3)_{color}$, and yet the interaction is not chiral. We believe this reasoning could offer a possible resolution of the strong CP problem. 

     There also appears to be an intriguing possibility that this gauge-gravity duality is also a bulk-boundary correspondence. Because Connes time is external to the octonionic space, one could think of the velocity $\dot q_B$ as having one extra dimension (hence bulk, gravity) compared to  the configuration variable $q_B$ (hence boundary, conformal field ?).

\bigskip

             5.\ \textbf{W and Z Bosons}:

             The subtlety about $SU(2)$ symmetries obeyed by the $W_1$, $W_2$, and $W_3$ bosons before symmetry breaking, and the breaking of the $SU(2)_L \times U(1)_Y$ into $U(1)_{em}$, is rather interesting. In our Lagrangian, we have representations of the photon, and the $W^+$, $W^-$ and the $Z^0$ bosons, i.e. post-symmetry breaking terms. It might appear surprising that the Lagrangian fundamentally has terms for the photon and the massive $W$ bosons, and not the massless electroweak bosons. However, this is consistent with the fact that the electroweak symmetry breaking in this theory coincides with the $E_8 \times E_8$ symmetry breaking. Prior to this transition, the Lagrangian is of a very different kind, i.e. the one that we started with, in Eqn. (\ref{acnbasis}). Hence it is apparent that in our theory, the formalism prior to the electroweak breaking departs from what one might expect from the standard model.

             Let us carefully examine how  the terms in Eqn. (\ref{weakgrav}) are a manifestation of the representations of the $W_1$, $W_2$, and $W_3$ bosons, and how these, in turn, are mappings of generators of $SU(2)$ symmetry. We now analyse the  $SU(2)$ symmetries of the weak bosons and their right-hand analogues - the Lorentz bosons. Intuitively, the $SU(2)_L$ weak symmetry and the $SU(2)_R$ pre-gravitation symmetry have resulted from left-right symmetry breaking so they must come from the mixed terms in the Lagrangian. In other words, we have $SU(3)\times SU(3)\longrightarrow SU(2)_L \times U(1)_Y \times SU(2)_R \times U(1)_g$ and even though the first $SU(3)$ comes from the left-handed sector and the second $SU(3)$ from the right handed sector, this branching mixes both the sectors. This is further evidence that the weak force and gravitation described by general relativity are connected to each other.
\vspace{0.5cm}

The three  bosons $W_1, W_2, W_3$ associated with the $SU(2)_L$ symmetry will be obtained from the following three $SU(2)_L$ generators as listed in \cite{f1, Vatsalya1}:
\begin{align} 
T_1= \tau_1 \frac{(1+ie_1 e_2)}{2}\; ;\qquad 
T_2= \tau_2 \frac{(1+ie_1 e_2)}{2}\; ; \qquad
T_3= \tau_3 \frac{(1+ie_1 e_2)}{2}
\end{align}
where \begin{align}  
\tau_1 = \Omega +\Omega ^\dagger\; ; \quad
\tau_2 = i\Omega -i\Omega ^\dagger\; ; \quad
\tau_3=\Omega \Omega^\dagger - \Omega ^\dagger \Omega \; ; \quad
\Omega= \alpha_1 \alpha_2 \alpha_3
\end{align}
and the expressions for the left-handed set of ladder operators $\alpha_i$ (i=1,2,3) are \cite{f1, f2, Vatsalya1, vvs, Singhfsc}:
\begin{align} 
\alpha_1=\frac{-e_5+ie_4}{2}\ \; \qquad
\alpha_2=\frac{-e_3+ie_1}{2}\ \; \qquad
\alpha_3=\frac{-e_6+ie_2}{2}
\end{align}
And for the RH set (the pre-gravitation Lorentz bosons), we have:
\begin{align} 
\alpha_1=-\omega\frac{-e_5+ie_4}{2}\ \; \qquad
\alpha_2=-\omega\frac{-e_3+ie_1}{2} \ \; \qquad
\alpha_3=-\omega\frac{-e_6+ie_2}{2}
\end{align}
Also, as is well known the relations between the pre-symmetry breaking bosons $(W_1, W_2)$ and the post-symmetry breaking bosons $(W^+, W^-)$ are given by:
\begin{align} 
W_1=\frac{1}{\sqrt{2}}(W^++W^-)\ \; \qquad
W_2=\frac{i}{\sqrt{2}}(W^+-W^-)
\label{dublew}
\end{align}
The analogous right-handed relations would be as follows:
\begin{align} 
W_{1R}=\frac{1}{\sqrt{2}}(W^+_R+W^-_R)\ \; \qquad
W_{2R}=\frac{i}{\sqrt{2}}(W^+_R-W^-_R)
\label{dublewr}
\end{align}

Let us suppress the factor of ${1}/{\sqrt{2}}$ in the calculations that proceed later. We will set up the transformation from the photon and $Z^0$ to $W_3$ and the weak hypercharge field $B$ when we talk of the weak mixing angle below.

We note here that the imaginary unit $i$ is now being used in the generators, whereas the octonionic space has been made using real, compact $E_8$ along with the octonions. However, the Grassmann number entries in the dynamical matrices are over the complex number field, so that effectively we now have complex split bioctonions. Complex octonions are essential also for constructing fermionic spinor states via the Clifford algebra $Cl(6)$ made from octonionic chains \cite{f1, f2}. Mention must also be made of the long history of fascinating research relating elementary particles and the standard model to the algebra of the octonions \cite{Dixon, Gursey, f1, f2, f3, Chisholm, Trayling, Dubois_Violette_2016, Todorov:2019hlc, Dubois-Violette:2018wgs, Todorov:2018yvi, Todorov:2020zae, ablamoowicz, baez2001octonions, Baez_2011,  baez2009algebra, f1, f2, f3, Perelman,  Gillard2019, Stoica, Yokota, Dray1, Dray2, lisi2007exceptionally, Ramond1976}.

Complexified octonions are needed also for recovering the emergent 4D classical spacetime through spontaneous localisation of highly entangled degrees of freedom \cite{Singhreview}.

\bigskip

\textbf{The Weak Bosons}: $\mathbf{SU(2)_L}$

The generators for $SU(2)_L$ symmetry (after explicit calculations following the above definitions) are given by:   
\begin{align}
T_1=-e_8 \frac{(1+ie_1 e_2)}{2} = \frac{1}{2} [-e_8(1.)-ie_8(e_4.)]\\ \nonumber
T_2=-1 \frac{(1+ie_1 e_2)}{2} =\frac{1}{2} [-1(1.)-i1(e_4.)] \\ \nonumber
T_3=ie_8 \frac{(1+ie_1 e_2)}{2} = \frac{1}{2}[ie_8(1.)-e_8(e_4.)]
\end{align}
We note that $T_3=-iT_1$ and, very significantly, that $T_2$ has a real direction (no $e_8$; this latter fact is what enables us to determine the weak mixing angle).
Let us designate the following mapping: (here, $e_k=e_ie_j$)
\begin{align}
& e_i(e_j.) \rightarrow (q_{Bi}^\dagger \dot{q}_{Bj} - q_{Bj}^\dagger \dot{q}_{Bi})\omega e_k \nonumber\\ 
& 1(1.) \rightarrow (-q_{B0} \dot{q}_{B0}+q_{B1}^\dagger \dot{q}_{B1}+q_{B2}^\dagger \dot{q}_{B2}+q_{B3}^\dagger \dot{q}_{B3}+q_{B4}^\dagger \dot{q}_{B4}+q_{B5}^\dagger \dot{q}_{B5}+q_{B6}^\dagger \dot{q}_{B6}+q_{B8}^\dagger \dot{q}_{B8})
\end{align}
Then, we get the mapped $T_i$ as follows:
\begin{align} 
& T_1\rightarrow [-(q_{B8}^\dagger \dot{q}_{B0}-q_{B0} \dot{q}_{B8})\omega e_8-(q_{B8}^\dagger \dot{q}_{B4}-q_{B4}^\dagger \dot{q}_{B8})\omega e_5] \nonumber \\
& T_2\rightarrow [ -  (-q_{B0} \dot{q}_{B0}+q_{B1}^\dagger \dot{q}_{B1}+q_{B2}^\dagger \dot{q}_{B2}+q_{B3}^\dagger \dot{q}_{B3}+q_{B4}^\dagger \dot{q}_{B4}+q_{B5}^\dagger \dot{q}_{B5}+q_{B6}^\dagger \dot{q}_{B6}+q_{B8}^\dagger \dot{q}_{B8})- \nonumber\\ & i(q_{B0}\dot{q}_{B4}-q_{B4}^\dagger \dot{q}_{B0})\omega e_4] \nonumber \\ 
& T_3 \rightarrow [i(q_{B8}^\dagger \dot{q}_{B0}-q_{B0} \dot{q}_{B8})\omega e_8-(q_{B8}^\dagger \dot{q}_{B4}-q_{B4}^\dagger \dot{q}_{B8})\omega e_5]
\label{TTT}
\end{align}
We will also assume, as is plausible, that these three bosonic terms are respectively proportional to $T_i^2$, the square of the weak isospin component  along that chosen direction. This is analogous to the bosonic term for the photon being proportional to the fine structure constant (square of electric charge).
Let us focus on the Lagrangian (\ref{weakgrav}) for a moment and assign the following interpretations to various terms therein (proportionality factor  $-iL_p^2 \alpha /L^3$ is suppressed for now).
\begin{align} \nonumber
-iW^+ \rightarrow i[(q_{B8}^\dagger \dot{q}_{B0}-q_{B0} \dot{q}_{B8}) +(q_{B3}^\dagger \dot{q}_{B1}-q_{B1}^\dagger \dot{q}_{B3})+(q_{B6}^\dagger \dot{q}_{B2}-q_{B2}^\dagger \dot{q}_{B6}) +(q_{B5}^\dagger \dot{q}_{B4} - q_{B4}^\dagger \dot{q}_{B5})] \omega e_8\\ \nonumber 
W^- \rightarrow i[(q_{B4}^\dagger \dot{q}_{B8}-q_{B8}^\dagger \dot{q}_{B4})+(q_{B1}^\dagger \dot{q}_{B6}-q_{B6}^\dagger \dot{q}_{B1})+(q_{B3}^\dagger \dot{q}_{B2}-q_{B2}^\dagger \dot{q}_{B3})+(q_{B5}^\dagger \dot{q}_{B0}-q_{B0} \dot{q}_{B5})]\omega e_5\\  
Z^0/\cos(\theta_W/2) \rightarrow i[(q_{B4}^\dagger \dot{q}_{B0}-q_{B0} \dot{q}_{B4})+(q_{B8}^\dagger \dot{q}_{B5}-q_{B5}^\dagger \dot{q}_{B8})+(q_{B3}^\dagger \dot{q}_{B6}-q_{B6}^\dagger \dot{q}_{B3})+(q_{B2}^\dagger \dot{q}_{B1}-q_{B1}^\dagger \dot{q}_{B2})]\omega e_4
\label{dubus}
\end{align}
The angle $\theta_W$ introduced here will subsequently be identified with the weak mixing angle.
Putting these into Eqn. (\ref{dublew}), we get the expressions for the pre-symmetry breaking $SU(2)_L$ bosons as follows:
\begin{align}
W_1=[-(q_{B8}^\dagger \dot{q}_{B0} - q_{B0} \dot{q}_{B8}) + ...]\omega e_8 +[-(q_{B8}^\dagger \dot{q}_{B4} - q_{B4}^\dagger \dot{q}_{B8})+ ...]i\omega e_5 \\ \nonumber
W_2=[ -(q_{B8}^\dagger \dot{q}_{B0} - q_{B0} \dot{q}_{B8}) + ...]i\omega e_8 +[-(q_{B8}^\dagger \dot{q}_{B4} - q_{B4}^\dagger \dot{q}_{B8}) + ...]\omega e_5
\end{align}
The three dots represent the additional terms which can be easily read off from the expressions in Eqn. (\ref{dubus}).
These correspond to the $SU(2)_L$ generators as follows, as can be easily verified by comparing with (\ref{TTT}).
\begin{align} 
W_1\rightarrow T_1  \ ; \qquad 
(W_2)^\dagger\rightarrow T_3 \ ; \qquad
W_3 \rightarrow T_2
\label{wgens}
\end{align}    
Therefore, we have shown that there are terms in (\ref{weakgrav}) that possess $SU(2)_L$ symmetry. The weak hypercharge field $B$ is defined as follows
\begin{equation}
B\; \tan(\theta_W/2) =  (-q_{B0} \dot{q}_{B0}+q_{B1}^\dagger \dot{q}_{B1}+q_{B2}^\dagger \dot{q}_{B2}+q_{B3}^\dagger \dot{q}_{B3}+q_{B4}^\dagger \dot{q}_{B4}+q_{B5}^\dagger \dot{q}_{B5}+q_{B6}^\dagger \dot{q}_{B6}+q_{B8}^\dagger \dot{q}_{B8})
\label{Bfield}
\end{equation} 
Given these definitions of $B$ and $Z^0$ and $W_3$ we find that
\begin{equation}
W_3 = Z^0 /\cos (\theta_{W}/2) + B \tan (\theta_{W}/2) 
\label{dub3}
\end{equation}


\bigskip

\textbf{The Lorentz bosons}: $\mathbf{SU(2)_R}$ \textbf{pre-gravitational symmetry:}

\smallskip

In the same way as for the left-handed case, we explicitly calculate the generators of $SU(2)_R$ symmetry:
\begin{align} \nonumber
T_1=\omega e_8 \frac{(1+ie_1 e_2)}{2} = \frac{1}{2} [\omega e_8 (1.)+i\omega e_8 (e_4.)] \\ \nonumber
T_2=\omega \frac{(1+ie_1 e_2)}{2} = \frac{1}{2} [\omega (1.)+i\omega (e_4.)] \\ \nonumber
T_3=ie_8 \frac{(1+ie_1 e_2)}{2} = \frac{1}{2}[ie_8(1.)-e_8(e_4.)]
\end{align}

\vspace{0.4cm}
Following the same procedure as before, let us designate the following (slightly different) mapping: (here, $e_k=e_ie_j$)

\begin{align}
& \omega e_i(e_j.) \rightarrow (\dot{q}_{Bi}^\dagger q_{Bj} - \dot{q}_{Bj}^\dagger q_{Bi})\omega e_k \nonumber \\ 
&  \omega(1.) \rightarrow (\dot{q}_{B0}q_{B0}-\dot{q}_{B1}^\dagger q_{B1} -\dot{q}_{B2}^\dagger q_{B2}-\dot{q}_{B3}^\dagger q_{B3}-\dot{q}_{B4}^\dagger q_{B4}-\dot{q}_{B5}^\dagger q_{B5}-\dot{q}_{B6}^\dagger q_{B6}-\dot{q}_{B8}^\dagger q_{B8})
\end{align}

 And for $T_3$, we use the mapping:
 \begin{equation} 
 e_i(e_j.) \rightarrow (\dot{q}_{Bi}^\dagger q_{Bj} - \dot{q}_{Bj}^\dagger q_{Bi})\omega e_k  
   \end{equation}

 We get the mapped $T_i's$ as follows:
 \begin{align}
 & T_1 \rightarrow [(\dot{q}_{B8}^\dagger q_{B0} - \dot{q}_{B0} q_{B8})\omega e_8 +i(\dot{q}_{B8}^\dagger q_{B4}-q_{B4}^\dagger q_{B8})\omega e_5]\nonumber \\ 
 & T_2 \rightarrow [(\dot{q}_{B0}q_{B0}-\dot{q}_{B1}^\dagger q_{B1} -\dot{q}_{B2}^\dagger q_{B2}-\dot{q}_{B3}^\dagger q_{B3}-\dot{q}_{B4}^\dagger q_{B4}-\dot{q}_{B5}^\dagger q_{B5}-\dot{q}_{B6}^\dagger q_{B6}-\dot{q}_{B8}^\dagger q_{B8})+i(\dot{q}_{B0} q_{B4}-\dot{q}_{B4}^\dagger q_{B0})\omega e_4] \nonumber \\ 
 & T_3 \rightarrow [i(q_{B8}^\dagger \dot{q}_{B0}-q_{B0} \dot{q}_{B8})\omega e_8-(q_{B8}^\dagger \dot{q}_{B4}-q_{B4}^\dagger \dot{q}_{B8})\omega e_5]
\end{align}


We assign some terms from the Lagrangian (\ref{weakgrav}) to the RH pre-gravitation bosons:
\begin{align} 
& -iW^{+}_R \rightarrow i[(\dot{q}_{B8}^\dagger q_{B0}-\dot{q}_{B0} q_{B8}) +(\dot{q}_{B1}^\dagger q_{B3}-\dot{q}_{B3}^\dagger q_{B1})+(\dot{q}_{B2}^\dagger q_{B6}-\dot{q}_{B6}^\dagger q_{B2}) +(\dot{q}_{B4}^\dagger q_{B5} - \dot{q}_{B5}^\dagger q_{B4})] \omega e_8 \nonumber \\ 
& W^-_R \rightarrow i[(\dot{q}_{B8}^\dagger q_{B4}-\dot{q}_{B4}^\dagger q_{B8})+(\dot{q}_{B6}^\dagger q_{B1}-\dot{q}_{B1}^\dagger q_{B6})+(\dot{q}_{B2}^\dagger q_{B3}-\dot{q}_{B3}^\dagger q_{B2})+(\dot{q}_{B5}^\dagger q_{B0}-\dot{q}_{B0} q_{B5})]\omega e_5 \nonumber \\ 
& Z^0_R/\cos (\theta_{g}/2)\rightarrow i[(\dot{q}_{B4}^\dagger q_{B0}-\dot{q}_{B0} q_{B4})+(\dot{q}_{B5}^\dagger q_{B8}-dot{q}_{B8}^\dagger q_{B5})+(\dot{q}_{B6}^\dagger q_{B3}-\dot{q}_{B3}^\dagger q_{B6})+(\dot{q}_{B1}^\dagger q_{B2}-\dot{q}_{B2}^\dagger q_{B1})]\omega e_4
\end{align}

Again substituting these into the expressions for what we call $W_{1R}$ and $W_{2R}$, we get,
\begin{align} 
& W_{1R}=[-(\dot{q}_{B8}^\dagger q_{B0}-\dot{q}_{B0} q_{B8})+...]\omega e_8 +[(\dot{q}_{B8}^\dagger q_{B4}-\dot{q}_{B4}^\dagger q_{B8})+...]i\omega e_5 \nonumber \\ 
& W_{2R}=[-(\dot{q}_{B8}^\dagger q_{B0}-\dot{q}_{B0} q_{B8})+...]i\omega e_8+[(\dot{q}_{B8}^\dagger q_{B4}-\dot{q}_{B4}^\dagger q_{B8})+...]\omega e_5
\end{align}

 These correspond to the $SU(2)_R$ generators in the following way:
 \begin{align} 
  -W^{\dagger}_{1R} \rightarrow T_1 \ ; \qquad 
  -W_{2R} \rightarrow T_3 \ ; \qquad
  W_{3R} \rightarrow T_2 
  \label{wgens2}
 \end{align}

 \vspace{0.5cm}
 There are some noteworthy points in the analysis given here. Firstly, as can be very clearly seen, we accounted for a small group of terms from the overall mixed part of the bosonic Lagrangian. The other terms can be easily obtained by simply relabelling the octonionic directions in the generators of $SU(2)$ symmetry. Physically, we can attribute this to gauge redundancy.
 Secondly, we have added and subtracted an extra total time derivative term $-\partial{_t} (q_{B0}q_{Bi})$ for each octonionic direction. We would thus have a leftover term:
 \begin{equation} \label{derivative}
 \partial{_t} (q_{B0}q_{B1}) +\partial{_t} (q_{B0}q_{B2})+\partial{_t} (q_{B0}q_{B3})+\partial{_t} (q_{B0}q_{B4})+\partial{_t} (q_{B0}q_{B5})+\partial{_t} (q_{B0}q_{B6})+\partial{_t} (q_{B0}q_{B8})
 \end{equation}
 for the mixed part of the bosonic Lagrangian. The significance of the leftover terms given by (\ref{leftoutgluon}), (\ref{leftoutgravigluon}) and (\ref{derivative}) will be investigated in future work.
 
It can be shown \cite{Singhspin1} that the bosonic degrees of freedom are all spin one, in the emergent quantum theory. Thus there is no elementary spin two graviton in this theory. Pre-gravitation is mediated by spin one gauge bosons associated with the $SU(2)_R\times U(1)_g$ symmetry. The graviton, if it exists, will be a composite of these spin one bosons. The point is that we do not quantise the metric, this being an emergent classical concept. What is quantised are the eigenvalues of the Dirac operator \cite{Landi1999, Rovelli} - these are in the nature of the square-root of the metric, and are naturally associated with  spin one bosonic degrees of freedom.

It can also be shown using the electric charge operator, i.e. the number operator which is associated with the $U(1)_{em}$ symmetry, that $W^+$ and $W^-$ have electric charge $+1$ and $-1$ respectively \cite{Vatsalya1} and that $Z^0$ is electrically neutral. The corresponding situation for the  $SU(2)_R \times U(1)_g$ symmetry is interesting, because here the $U(1)_g$ number operator defines square root of mass (in Planck mass units); it does not define  electric charge. Consequently, $W^+_R$ and $W^-_R$ have square root mass $+1$ and $-1$ respectively, and hence their range of interaction is limited to Planck length. They will also have an extremely tiny electric charge, some seventeen orders of magnitude smaller than the charge of the electron (analogous to the $W$ mass being so small on the Planck scale). Whereas the $U(1)_g$ boson (and the gravi-photon it transforms to) will have zero mass and zero electric charge. $Z^0_R$ will be massless, and will have  an extremely tiny electric charge (like the $W_R$ bosons). It is possible that emergent gravitation is mediated at the quantum level by the $Z^0_R$ and the gravi-photon. They take the place of the spin 2 graviton, in this theory. It is possible that these two are the gauge bosons corresponding to the Dirac eigenvalues which Landi and Rovelli have proposed as the dynamical variables for general relativity \cite{Rovelli}, in place of the metric. This possibility will be investigated further, so as to understand how this could imply the emergence of general relativity, perhaps with a modification at galactic and cosmic length scales.

\subsection {Theoretical Derivation of the numerical value of the weak mixing  angle}


Spontaneous symmetry breaking makes the $W_3$ and $B$ bosons transform into two different physical bosons with different masses - the massive $Z^0$ boson and the massless photon ($\gamma$). It can be viewed as a rotation in the gauge field space defined by the octonions. Let the rotation angle be ${\theta_W}/{2}$. Here $\theta_W$ is the weak mixing angle. We have taken the rotation angle as ${\theta_W}/{2}$ because this rotation is taking place on the spinorial equivalent of the conventional gauge field space related to Minkowski spacetime. Rotation by $\theta_W/2$ on octonionic space corresponds to a rotation $\theta_W$ in conventional gauge space.

\begin{equation}
\begin{pmatrix}
    \gamma \\
    Z^0
\end{pmatrix}
=
\begin{pmatrix}
    \cos (\theta_W/2) & \sin (\theta_W/2)\\
    -\sin (\theta_W/2) & \cos (\theta_W/2) \\
\end{pmatrix}
\begin{pmatrix}
    B \\
    W_3
\end{pmatrix}
\end{equation}

The expressions for B and $W_3$ are as follows:

\begin{align}
B=  \cos (\theta_W/2) \gamma - \sin (\theta_W/2) Z^0\\
W_3 =\sin (\theta_W/2) \gamma + \cos (\theta_W/2) Z^0 
\end{align}
If we relate the photon term $\gamma$ given in (\ref{photon}) and the $Z^0$ (Eqn. (\ref{dubus}) from the Lagrangian to the $W_3$ defined in (\ref{wgens}) using the above rotation matrix, then clearly the Lagrangian obeys an $SU(2)_L \times U(1)_Y$ symmetry. Moreover,  the weak hypercharge $B$ with $U(1)_Y$ symmetry defined  by the above rotation is also equated to the expression given in (\ref{Bfield}). 

We can relate the expression for  $\gamma$  given by (\ref{photon}) to the result from the above rotation matrix by first expressing $\gamma$ from the matrix in terms of 
$Z^0$ and $B$ from Eqns. (\ref{dubus}) and (\ref{Bfield}) respectively, and then identifying this $\gamma$ with the one in (\ref{photon}):
 \begin{equation}
\gamma = \frac{B}{\cos(\theta_W/2)} + Z^0 \tan(\theta_W/2)
\label{gammay}
 \end{equation}

 The expression for $W_3$ given in (\ref{dub3}) in terms of $Z^0$ and $B$ using the Lagrangian, is the same relation as obtained from the above rotation matrix. Furthermore, when we substitute for $Z^0$ from (\ref{dub3}) in (\ref{gammay}) above, we get the relation between $\gamma, B$ and $W_3$ precisely as required by the rotation matrix.
 \begin{equation}
 \gamma= {B}{\cos(\theta_W/2)}  + W_3 \sin(\theta_W/2)
 \end{equation}
 Hence there is overall consistency between the post-SSB Lagrangian and the pre-SSB Lagrangian.

While deriving the asymptotic mixing angle, we will be taking the low energy limit, $q_{Bi} \rightarrow kL_p$, where k is such that the expressions defined by $q_B$ and $\dot{q}_B$ are normalised to unity. This is the asymptotic `flat spacetime' limit in which the interactions determined by $q_B$ are not yet switched on. In this limit the $Z^0$, being antisymmetric, goes to zero. Therefore, we can drop it from the expressions for $B$ and $W_3$ above, which now depend only on $\gamma$.
\begin{align}
B= \frac{L_p^4}{L^4} \alpha ^2\cos (\theta_W/2)=Q^2\cos (\theta_W/2)\nonumber\\
W_3= \frac{L_p^4}{L^4} \alpha ^2\sin (\theta_W/2)=Q^2\sin (\theta_W/2)
\label{a}
\end{align}
      where  $Q^2={\alpha^2 L_p^4}/{L^4}  $ is the fine structure constant \cite{Singhfsc}.


It is also well known that the electric charge $Q$ is a non- trivial combination of the $Y_W$ (weak hypercharge) and the $T_3$ component of weak isospin as follows:

\begin{equation} Q=\frac{Y_W}{2} + T_3 \label{hypQ} \end{equation}
We will make the plausible assumptions that the $B$ field is proportional to the square of the hypercharge $Y_W$ and the field $W_3$ is proportional to the square of $T_3$, i.e. $T_3^2$. Therefore we get the limiting expressions for $B$ and $W_3$ to be $Y_W^2$ and $T_3^2$ respectively, giving from (\ref{a}) the relations
\begin{align}
Y_W^2 \rightarrow B \implies Y_W \rightarrow Q \sqrt{\cos(\theta_W/2)}\\
T_3^2 \rightarrow W_3 \implies T_3 \rightarrow Q\sqrt{\sin(\theta_W/2)} \end{align}


Substituting these into the expression  (\ref{hypQ}) for the electric charge $Q$ gives us the trigonometric equation:


\begin{equation} 1= \frac{\sqrt{\cos(\theta_W/2)}}{2} + \sqrt{\sin(\theta_W/2)} \label{telltheta} \end{equation}


Apart from $\theta_W=\pi$ the other solution to this equation is (see Fig. 3) :





\begin{equation}\theta_W = 29.98^{\circ} \qquad
\implies \sin^2 \theta_W = 0.2496977...  \end{equation}

\begin{figure}[ht]
    \includegraphics[width = 8cm]{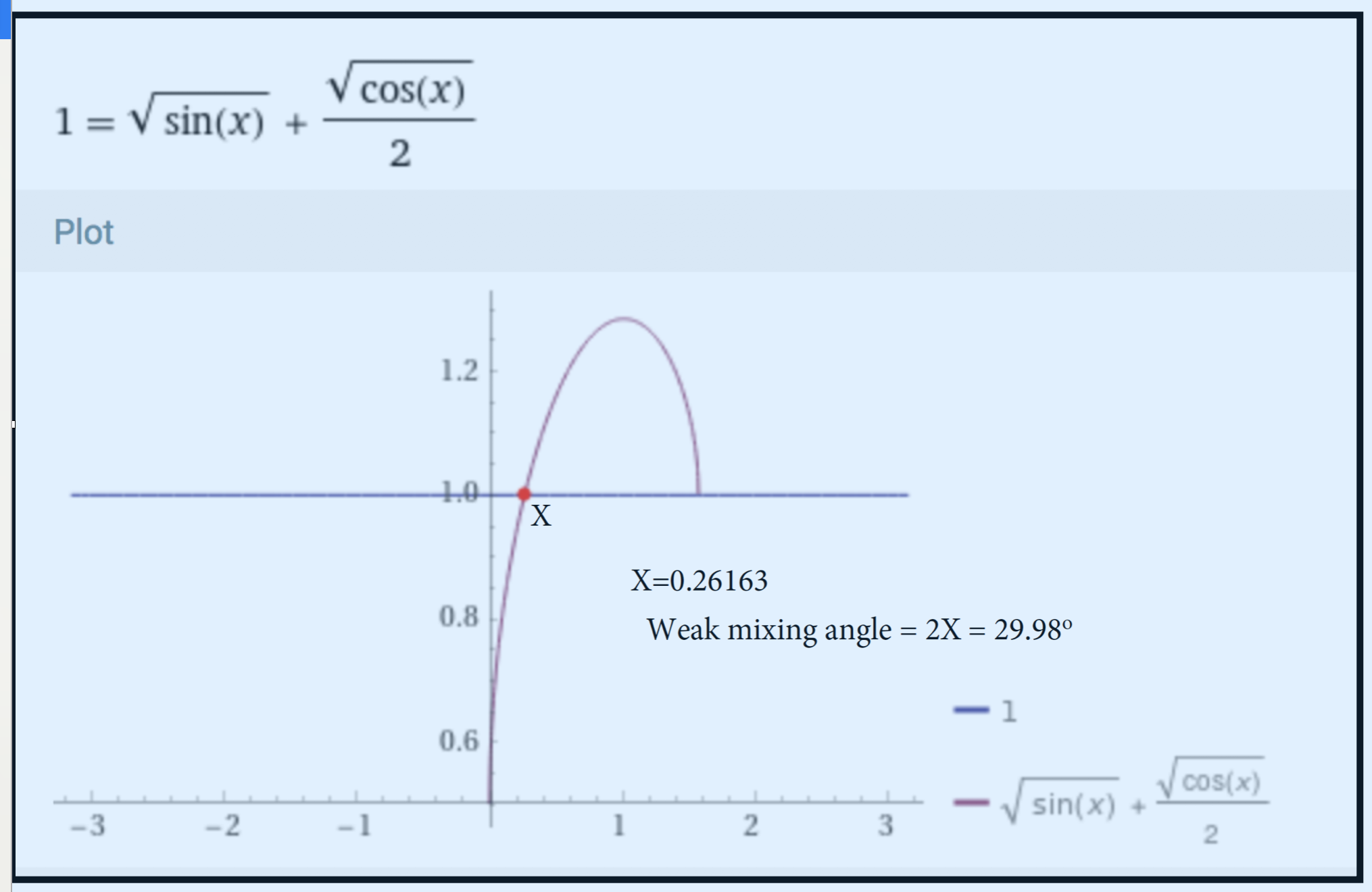}
    \caption{Solution to the governing equation for the weak mixing angle}
\end{figure}
We note that we have taken the limit $q_B\rightarrow L_p$ as the `flat limit'. What limit should have been assigned to $\dot{q}_B$ in the expressions for $B$ and $W_3$, which we note have an overall coefficient $-iL_P^2 \alpha /L^3$ (from Eqn. (17))? We have assumed  $\dot{q}_B$ to go to a constant in the flat limit where $q_B$ goes to $L_p$. There is no guarantee a priori that this constant  will be unity; nor that it will be the same for the $B$ field and for $W_3$. Hence we have combined the constant value with  $L_p^3 \alpha /L^3$ and defined the new constants  to be $Y^2$ and 
$T_3^2$ respectively, which is completely plausible.

The CODATA 2018 experimental value for the mixing angle is given by $\sin^2{\theta_{W}} = 0.22290(30)$ which is smaller than the result we find above. However, the value we find is the asymptotic no-interaction limit and it remains to be found out exactly how the value we find runs with energy, so that we may then match theory  against the measured value, which is typically measured  around 90 GeV scale.

One could possibly arrive at an equation identical to Eqn. (\ref{telltheta}) also in the QFT treatment of the standard model, but with the angle in the equation as $\theta_W$ but not half of $\theta_W$. Without that half, this equation will yield $\sin^2\theta_W = 0.0669$ which strongly disagrees with the measured value. This could be taken as further evidence that the spinor nature of spacetime is crucial for determing standard model parameter values, as seen also in the case of the fine structure constant and mass ratios.

\section{Outlook}
In future work. we plan to investigate the interaction terms in the original Lagrangian above, which describe the action of the bosons on the fermions. In particular, it is important to establish that three chiral fermion generations are present, whereas the bosons must not to be triplicated. Some preliminary effort on this aspect has been made in \cite{Singhfsc}. Subsequently, the bifermion terms need to be looked at, to understood their possible  role  in the Higgs mechanism, and whether they help account for the unaccounted for degrees of freedom reported in \cite{priyank}. The eventual goal of course is to see if this elementary Lagrangian describes the standard model and general relativity, while also prediciting some beyond standard model new physics.

The left-right symmetry breaking could also help understand the origin of matter-antimatter asymmetry, as has been explained in \cite{singhessay} (please see Section V therein). 

In this theory, quarks live in eight dimensions, i.e. the octonionic space. It is in the very nature of their definition. It is impossible to confine a quark in 4D spacetime; and since all our measurements take place in the emergent 4D spacetime, such measurements will never detect an 8D object such as an isolated quark. The only way to detect quarks is as composite states which have zero color, and since color in this theory is entirely  a geometric feature, zero color is the same thing as saying the composite state is effectively in our 4D spacetime.
     
Leptons, on the other hand, can be described on quaternionic space; more precisely, split biquaternionic space, if we are to take into account their chiral nature \cite{Vatsalya1}. This space is associated with the Clifford algebra $Cl(3)$ and with the $SU(2)_L\times SU(2)_R$ symmetry of the leptons of this theory. The associated spacetime is likely four-dimensional, with a LH $SO(4)$ counterpart. Since the weak force in this theory is more like gravitation than like an internal symmetry, the L-R symmetry breaking is happening in the gravitational sector, but not in the gauge sector $SU(3)_c \times SU(3)_{grav}$. The breaking could perhaps be understood as a result of separation of matter and antimatter. The former has positive square-root mass and the latter has negative square root mass; and matter and antimatter repel gravitationally. In the very early unverse, an octonionic scale-invariant inflationary phase (having $E_8\times E_8$ symmetry) ends with the quantum-to-classical phase transition, which forms classical matter as well as classical antimatter. Gravitational repulsion separates matter from antimatter faster than they can annihilate, resulting in our matter only universe in which gravitation appears only as an attractive interaction (matter-matter attracts, antimatter-antimatter attracts, matter-antimatter repels, gravitationally). Our part of the universe violates CP as well as T, and by itself the (backward moving in time) antimatter universe also violates CP as well as T, but together the two universes preserve CPT (see also \cite{BoyleTurok}). The scale invarinace of the original universe is replaced by CPT symmetry of the emergent matter-antimatter universe.

We predict three right handed sterile neutrinos, one per generation, and moreover the neutrinos in this theory are Majorana fermions. The neutrino in this theory is likely massless, with flavor oscillations possibly explained geometrically, because neutrinos in this theory are spacetime triplets, the only particle to have that property, irrespective of parity \cite{priyank}.
     
\noindent{\it The Dirac operator} : In one of his lectures, Atiyah reminds us \cite{atiyah}  that the Dirac operator was actually first discovered by Hamilton when he found the quaternions. Because, given the three imaginary directions $\hat{i}, \hat{j}, \hat{k}$ of a quaternion, one can construct the gradient operator $D$ on 3-d space
\begin{equation}
D = \hat{i} \frac{d\;}{dx} + \hat{j} \frac{d\;}{dy} + \hat {k} \frac{d\;}{dz}     
\end{equation}    
       whose square is clearly the Laplace operator. One can include also the real direction of the quaternion in the  gradient operator, and then the square of the gradient operator is the d'Lambertian on 4D Euclidean space. By using a split quaternion instead of a quaternion, the gradient operator can be squared to get the Klein-Gordon operator, and hence Minkowski spacetime; so that now the gradient operator is  the usual Dirac operator. The associated space on which the Dirac operator is the gradient is of course not Minkowski spacetime, but its square root, the split quaternionic space. From the complex split quaternion, one can make a Clifford  algebra, spinors made from which describe fermions. The quaternion and the corresponding Dirac operator form a canonical pair.

       On Minkowski spacetime, if one introduces the Riemannian geometry via a metric, and defines the usual Dirac operator on it, the eigenvalues of the Dirac operator play the role of dynamical variables of general relativity. The trace of the square of the Dirac operator, upon a regularised heat kernel expansion, is proportional to the Einstein-Hilbert action on the manifold. This is the famous spectral action principle \cite{Chams:1997}.

       Evidently, Minkowski space-time is not the same as the 4D split-quaternionic space, but rather its classical limit, in the following well-defined physical sense. Let us define  an `atom of space-time' by the action principle $S = \int\; d\tau D_s^2$, where $D_s$ is the Dirac operator on a split quaternion space, and $\tau$ is Connes time. Consider a universe made of a very large number of atoms of space-time, each having its own Dirac operator $D_{si}$. The action principle for the universe then is $\int \; d\tau\; \sum_i \; D_{si}^2$. For each of the spacetime atoms we have its own split quaternionic space, and its own Dirac operator, and we also associate with it the bosonic configuration variable $\dot{q}_{Bi}\equiv D_{si}$. Thus the action of the spacetime atom is $\int \;d\tau \;\dot{q}_B^2$ and this is how we make contact with the pre-quantum theory of trace dynamics. This action can be generalised to include Yang-Mills fields $q_B$, and fermions $q_F$ and $\dot{q}_F$, so as to have a generalised Dirac operator, thus having an action of the form (\ref{acnbasis}), an atom of space-time-matter (STM), or the aikyon. In general, the associated Hamiltonian has an anti-self-adjoint component which, during evolution,  causes spontaneous localisation to one or the eigenvalues $\lambda_i$ of the generalised Dirac operator (collapse of the wave function). This leads to a process of the following kind, when many STM atoms localise:

       \begin{equation}
\int \; d\tau \; \left[ D_1^2 + D_2^2 + D_3^2 + ... \right] \longrightarrow \int \; d\tau \left[ \lambda_1^2 + \lambda_2^2 + \lambda_3^2 + ...\right] = \int \; d\tau\; D^2 \propto \int \; d\tau \;  \int d^4x \; \sqrt{-g} \; [R + {\cal L}_{matter}]
\label{emergence}
       \end{equation}
The eigenvalues to which the different fermions localise serve to define  the coordinates of the emergent (curved) classical 4D spacetime. In this sense, the spontaneous localisation of a collection of many copies of the quaternionic space gives rise to classical matter on classical spacetime. Those STM atoms which have not undergone spontaneous localisation continue to be described, as before, by trace dynamics, or to an excellent approximation, by quantum field theory on the emergent classical spacetime.  

We see that whereas it is perfectly legitimate to construct the Dirac operator on Minkowski spacetime, doing so is an approximation to defining the Dirac operator on (split) quaternionic space; the latter being the natural home of the Dirac operator. This is yet another way of motivating the present approach to unification, supplementing what was stated at the very start of Section I.

Analogously, one can extend the idea of the Dirac operator to the split biquaternions, where the $SU(2)_L \times SU(2)_R$ symmetry is manifest, and we clearly see general relativity as the right-handed counterpart of the weak force.

In the same spirit, one defines the Dirac operator as the gradient operator on octonionic space and then on split bioctonionic space, leading to the Lagrangian in (\ref{acnbasis}). Here, $\dot{\widetilde Q}_1$ is the Dirac operator and so is $\dot{\widetilde Q}_2$. The consistency of the trace dynamics equations requires that $\beta_1 \ne \beta_2$, thus ensuring that one has a bilinear form describing a 2-brane on bioctonionic space, and not a quadratic form which is the Klein-Gordon operator on 10D Minkowski spacetime.  The branching proposed in \cite{priyank} and shown also at the beginning of this paper in (\ref{branching}) undoubtedly bears some relation to the Freudenthal-Tits magic square and also to the octonionic projective plane $OP^2$ and the groups $G_2, F_4, E_6, E_7$. These connections will be investigated in future work. Furthermore, these remarks on the Dirac operator help understand why the Dirac equation for three fermion generations emerges after solving the equations of motion.

 \bibliography{biblioqmtstorsion}


\end{document}